\documentclass{article}
\usepackage{diagrams}
\usepackage{amssymb}
\usepackage{verbatim}
\def\ds{\displaystyle}
\def\g{\gamma}

\def\l{\lambda}
\def\un{\underline}
\def\KK{{\mathbf K}}

\def\vrul{\rule[20pt]{0pt}{0pt}}

\newtheorem{theorem}{Theorem}[section]
\newtheorem{examp}{Example}[section]
\newtheorem{coroll}{Corollary}[section]
\newtheorem{examps}{Examples}[section]

\newtheorem{lemma}{Lemma}[section]
\newtheorem{remark}{Remark}[section]

\newtheorem{remarks}[remark]{Remarks}
\newtheorem{proposition}{Proposition}[section]
\newtheorem{definition}{Definition}[section]

\def\le{\left}
\def\m{\mathop}

\def\Bmat{\mathop{\mathbb  B}}

\def\Amat{\mathop{\mathbb A}}
\def\ri{\right}
\def\br{\begin{remark}\rm\small}
\def\1{{\bf 1}}
\def\er{\end{remark}}
\def\bt{\begin{theorem}\rm}
\def\et{\end{theorem}}
\def\bc{\begin{coroll}\rm}
\def\ec{\end{coroll}}
\def\brs{\begin{remarks}.\\ \rm\small\begin{enumerate}}
\def\ers{\end{enumerate}\end{remarks}}
\def\bx{\begin{examp}\small}
\def\ex{\end{examp}}
\def\bl{\begin{lemma}\small}
\def\el{\end{lemma}}
\def\bxs{\begin{examps}. \rm\begin{enumerate}}
\def\exs{\end{enumerate}\end{examps}}
\def\bd{\begin{definition}}
\def\ed{\end{definition}}
\def\bp{\begin{proposition}\rm}
\def\ep{\end{proposition}}
\def\be{\begin{equation}}
\def\ee{\end{equation}}

\def\bea{\begin{eqnarray}}
\def\eea{\end{eqnarray}}
\def\beas{\begin{eqnarray*}}
\def\eeas{\end{eqnarray*}}
\def \pa{\partial}
\def\C{{\mathbb C}}
\def\A{\mathop{\mathbf a}}
\def\B{\mathop{\mathbf b}}
\def\L{\Lambda}
\def\R{{\mathbb R}}
\def\N{{\mathbb N}}

\def\Z{{\mathbb Z}}
\def\a{{\alpha}}
\def\b{{\beta}}
\begin{document}
\begin{flushright}
CRM-2828 (2001)\\
\hfill Saclay-t01/123
\end{flushright}
\vspace{0.2cm}
\begin{center}
\begin{Large}
\textbf{Duality of spectral curves arising in two--matrix models}\footnote{
Based in part on a talk given by J. Harnad at the NEEDS 2001 Euroconference,
24--31 July 2001, Isaac Newton Institute for Mathematical Sciences,
Cambridge, U.K.}
\end{Large}\\
\vspace{1.0cm}
\begin{large} {M.
Bertola}$^{\dagger\ddagger}$\footnote{bertola@crm.umontreal.ca}, 
 { B. Eynard}$^{\dagger
\star}$\footnote{eynard@spht.saclay.cea.fr}
 and {J. Harnad}$^{\dagger \ddagger}$\footnote{harnad@crm.umontreal.ca}
\end{large}
\\
\bigskip
\begin{small}
$^{\dagger}$ {\em Centre de recherches math\'ematiques,
Universit\'e de Montr\'eal\\ C.~P.~6128, succ. centre ville, Montr\'eal,
Qu\'ebec, Canada H3C 3J7} \\
\smallskip
$^{\ddagger}$ {\em Department of Mathematics and
Statistics, Concordia University\\ 7141 Sherbrooke W., Montr\'eal, Qu\'ebec,
Canada H4B 1R6} \\ 
\smallskip
$^{\star}$ {\em Service de Physique Th\'eorique, CEA/Saclay \\ Orme des
Merisiers F-91191 Gif-sur-Yvette Cedex, FRANCE } \\
\end{small}
\bigskip
\bigskip
{\bf Abstract}
\end{center}
\begin{center}
\begin{small}
\parbox{13cm}{
The two matrix model is considered with measure given by the
exponential of a sum of polynomials in two different variables. It is
shown how to derive a sequence of pairs of ``dual'' finite size
systems of ODEs for the corresponding biorthonormal polynomials. An
inverse theorem is proved showing how to  reconstruct such measures
from pairs of semi-infinite finite band matrices defining the
recursion relations and satisfying the string equation.
A proof is given in the  $N\to \infty$ limit that the dual systems
obtained share the same spectral curve.
}
\end{small}
\end{center}
\baselineskip 14 pt

\section{Introduction}
We  consider the two--matrix model 
\cite{KazakoVDK,eynard,eynardchain,eynardmehta,McLaughlin}, which
 involves an ensemble consisting of pairs of 
$N\times N$ hermitian matrices $M_1$ and $M_2$, with a $U(N)$ invariant
probability measure of the  form: 
\be
{1 \over \tau_N}d\mu(M_1,M_2):= {1\over \tau_N} \exp{K {\rm tr}\, 
\left( -V_1(M_1) - V_2(M_2) + M_1 M_2 \right)} dM_1 dM_2  \ .
\label{twomatrixmeas}
\ee
Here $dM_1 dM_2$ is the standard Lebesgue measure for pairs of Hermitian 
matrices and $V_1$ and $V_2$ are chosen to be polynomials of degrees 
$d_1+1$, $d_2+1$ respectively, and are referred to as the potentials. The
overall positive scaling factor $K$ in the exponential is  taken as having
order $N$ when considering the large $N$ limit. We also assume  that both
potentials are real and bounded from below (for reasons of  convergence) .

 The normalization factor (partition function) 
\be
\label{deftau}
\tau_N = \int_{M_1}\int_{M_2} d\mu \ 
\ee
is known to be a KP $\tau$-function in each set of deformation
parameters (the coefficients of the two polynomials $V_1,V_2$),
 as well as providing solutions to the two-Toda equations
\cite{UT, AvM1, AvM2}.
The key objects of the theory are the correlation functions for the
eigenvalues of the two matrices.  Analogously to  the one--matrix models, 
such correlation functions can be recovered by means  of certain Fredholm 
integral  kernels.
We recall here briefly that in one-matrix models  with measure
\be
{1 \over \tau_N}d\mu(M):= {1\over \tau_N} \exp{ {\rm tr}\, 
\left( -V(M)  \right)} dM  
\ee
the computation of such a kernel  is reduced to
the construction of orthonormal polynomials $P_n(x)$ 
for the space
$L^2\le(\R, {\rm e}^{-V(x)}{\rm d}x\ri)$. 
In terms of these polynomials, the kernel is given by
\be
\m{K}^N(x,x') = \sum_{n=0}^{N-1} P_n(x){\rm e}^{-\frac 1 2 V(x)}P_n(x')
{\rm e}^{-\frac 12 V(x')}\ .
\ee
In  $2$-matrix models 
there are four 
relevant kernels needed to  compute  the statistical
correlations of eigenvalues. For $m$--matrix models there are $m^2$
such kernels.\par
These kernels are expressible in terms of suitably defined sequences of {\em
biorthogonal} polynomials.
By this  we mean two sequences of monic polynomials
\be
\pi_n(x) = x^n + \cdots , \qquad \sigma_n(y)=y^n + \cdots, \qquad n=0,1,\dots
\ee
which are orthogonal with respect to a coupled measure on the product space:
\be
\int_\R\!\!\int_\R\!\!\! {\rm d}x\, {\rm d}y \,\, \pi_n(x)\sigma_m(y) {\rm
e}^{-K V_1(x)-K V_2(y) +Kxy} = h_n\delta_{mn} ,
\ee 
where $V_1(x)$ and $V_2(y)$ are  the polynomials
appearing in the two-matrix model measure (\ref{twomatrixmeas}).
  The orthogonality 
relations determine the two families uniquely, if they exist \cite{McLaughlin}.
The four relevant kernels are expressed as follows in terms of these
biorthogonal polynomials 
\bea
&&{\m{K}^N}_{12}(x,y) =  \sum_{n=0}^{N-1} {1\over h_n} \pi_n(x) 
\sigma_n(y) {\rm  e}^{-KV_1(x)} {\rm  e}^{-KV_2(y)}  \ ,   
\\
&&{\m{K}^N}_{11}(x,x') = \int_\R {{\rm d}y \,\, {\m{K}^N}_{12}(x,y)\, {\rm 
e}^{Kx' y}}, \\ &&{\m{K}^N}_{22}(y',y) = \int_\R {{\rm d}x \,\,
{\m{K}^N}_{12}(x,y)\, {\rm  e}^{Kx y'}}  \ , 
\\ &&   {\m{K}^N}_{21}(y',x') =
\int_\R \int_\R {{\rm d}x
\,{\rm d}y  {\m{K}^N}_{12}(x,y)\, {\rm  e}^{Kxy'}{\rm  e}^{Kx'y}}  \ . 
\label{Ker12}
\eea
All the statistical properties of the spectra of the 2-matrix ensemble may
then be expressed in terms of these kernels \cite{eynardmehta} and the
corresponding Fredholm integral operators $\ds{{\m{\KK}^N}_{ij}}, \ i,j=1,2$.
For instance the density of eigenvalues of the first matrix is:
\be 
{\m{\rho}^N}_1(x) = {1\over N}\, {\m{K}^N}_{11}(x,x) \ ,
\ee
the correlation function of two eigenvalues of the first matrix is:
\be 
{\m{\rho}^N}_{11}(x,x') = {1\over N^2}\left({\m{K}^N}_{11}(x,x) {\m{K}^N}_{11}(x',x') 
- {\m{K}^N}_{11}(x,x') {\m{K}^N}_{11}(x',x)\right) \ ,
\ee
and the correlation function of two eigenvalues, one of the first matrix and one
of the second is:
\be
{\m{\rho}^N}_{12}(x,y) = {1\over N^2}
\left({\m{K}^N}_{11}(x,x) {\m{K}^N}_{22}(y,y) - 
{\m{K}^N}_{12}(x,y) ({\m{K}^N}_{21}(y,x) - e^{Kx y})\right) \ . 
\ee
Any other correlation function of $m$ eigenvalues can similarly be written 
as a determinant involving these four kernels.\par
The main objective of this paper is to derive and analyze certain
 differential systems of ODE's satisfied by the quasipolynomials
 $\psi_n(x) := \pi_n(x){\rm e}^{-V_1(x)}$, $\phi_n(y):=
 \sigma_n(y){\rm e}^{-V_2(y)}$ and their Fourier Laplace transforms.
In section \ref{due}, we summarize the principal results for
 finite $N$. The details and proofs may be found in  \cite{BEH} and \cite{BEHH}.
In section \ref{tre} we derive the corresponding results
 in the  $N\to\infty$  limit in a simple way. \par
The proof of Prop. \ref{jacqprop} and the non--abelian version of the
 transversality argument in Section \ref{tre} is based on joint work
 with J. Hurtubise, details of which will appear in \cite{BEHH}.
\section{Folding and systems of ODE in duality}
\label{due}
 Consider the normalized quasi-polynomials
\be
 \psi_n(x) = \frac 1{\sqrt{h_n}} \pi_n(x){\rm e}^{-KV_1(x)} \ , \qquad
\phi_n(y) = \frac 1{\sqrt{h_n}} \sigma_n(y) {\rm e}^{-KV_2(y)}\ , 
n=0, \dots \infty  \ .
\ee
Viewing these as components of a pair of semi--infinite
column vectors 
\be
\ds{\m{\Psi}_\infty}= (\psi_0, \psi_1, \dots \psi_n, \dots )^t \quad
{\rm and } \quad
 \ds{\m{\Phi}_\infty}=(\phi_0, \phi_1, \dots \phi_n, \dots )^t \ ,
\ee
 we obtain 
a pair of semi-infinite matrices $Q$ and $P$ that implement
multiplication of $\ds{\m{\Psi}_\infty}$ by $x$ and derivation
$-\frac 1K {d\over dx}$, respectively. Equivalently, we obtain the
transposes $Q^t$ and 
 $P^t$ by applying $-\frac 1K {d \over dy}$ or multiplication by $-y$ to 
$\ds{\m{\Phi}_\infty}$. By construction, these satisfy the Heisenberg 
commutation relations (or ``string equation'')
\be
[P, Q] = -\frac 1 K  \1 \ .
\ee
Along with these quasipolynomials we  need their
Fourier-Laplace transforms and the corresponding semi-infinite
(row)-vectors with components
\bea
&& \un\psi_n(y):= \int_\R{\rm d}x \,{\rm e}^{Kxy}\psi_n(x)\ ,\ \  \un\phi_n(x):=
\int_\R {\rm d}y \, {\rm e}^{Kxy}\phi_n(y)\\
&& \m{\un\Psi}_{\infty}(y) :=(\un\psi_0,...,\un\psi_n,...)\ ;\ \
\m{\un\Phi}_{\infty}(x) :=(\un\phi_0,...,\un\phi_n,...) \ .
\eea
The multiplicative and derivative recursion relations for these
sequences can be shown (by integration by parts) to be
\bea
&& x\m{\un\Phi}_\infty(x) = \m{\un\Phi}_\infty(x) Q\ ;\ \frac 1K \frac {d}{dx}
\m{\un\Phi}_\infty(x) = \m{\un\Phi}_\infty(x) P\\
&& y\m{\un\Psi}_\infty(y) = \m{\un\Psi}_\infty(y) Q^t\ ;\ \frac 1K \frac {d}{dy}
\m{\un\Psi}_\infty(y) = \m{\un\Psi}_\infty(y) P^t\ .
\eea
It also follows \cite{BEH} from  integration by parts that the two matrices $P$ and
$Q$ have a finite band structure
\bea
&& Q :=\le[
\begin{array}{ccccc}
 \a_0(0) &  \gamma(0) & 0 & 0 &\cdots \\
 \a_1(1) & \a_0(1) &\gamma(1) & 0 &\cdots \\
  \vdots &\hspace{-50pt}
\ddots
 &\hspace{-50pt}\ddots
 &  
\hspace{-50pt}\ddots
 & \hspace{-50pt} \ddots
\cr
\a_{d_2}(d_2) &\cdots & \a_0(d_2)& \gamma(d_2)& 
 \cr
0 &\hspace{-50pt}
\ddots&\hspace{-50pt}
\ddots&\hspace{-50pt}
\ddots&\hspace{-50pt}
\ddots
\end{array}  
\ri]\label{Qdef}\\
 && P := \le[
\begin{array}{ccccc}
 \b_0(0) &   \b_1(1) & \cdots & \b_{d_1}(d_1)& \cdots\\
\gamma(0) & \b_0(1) & \b_1(2)&\ddots& \b_{d_1}(d_1\!+\!1)\\
0 & \gamma(1) & \b_0(2) & \ddots &\!\!\!\!\!\!\!\!\!\!\!\!\!\!\!\!\ddots \\
0 &0&  \gamma(2) & \b_0(3) &\!\!\!\!\!\!\!\!\!\!\!\!\!\!\!\!\ddots\\
\vdots& \ddots & \ddots &\ddots &\!\!\!\!\!\!\!\!\!\!\!\!\!\!\!\! \ddots
\end{array}  
\ri],
\label{Pdef}
\eea
%
%
where $\gamma(n)\neq 0$ for all $n\in \N$.
This structure essentially follows from the fact that the two matrices
\be
(P-V_1'(Q)),\ \ (Q-V_2'(P))
\ee
are strictly lower and upper triangular respectively. Indeed, in the
basis of quasipolynomials it is obvious that
\bea
&&\sum_{m=0}^{\infty}(P-V_1'(Q))_{nm}\psi_m(x) =\le(-\frac 1K\frac
d{dx}-V_1'(x)\ri) \psi_n(x)
\cr &&  = c\psi_{n-1}(x) + \hbox{lower degree
quasipolynomials.}
\eea
and that $Q$ and $P^t$, representing the multiplication by $x$ and $y$
respectively, can have no more than one diagonal above the main
diagonal.
The converse is also true as will be detailed below.
\bp
\label{jacqprop}
Suppose that $P$ and $Q$ have the above band structure and that the
highest diagonal of $Q$ and the lowest of $P$ have nonzero entries.
Then the two following conditions are equivalent
\begin{enumerate} 
\item [$(i)$]
The commutator $[P,Q]$ is diagonal.
\item [$(ii)$] There exist two polynomials of degrees $d_1$ and $d_2$
respectively which we denote by
$V'_1(x)$ and $V_2'(y)$ such that 
\be
\le(P-V'_1(Q)\ri)_{\geq 0} = 0\ ,
\ \ 
\le(Q-V'_2(P)\ri)_{\leq 0} = 0\ , \label{triang}
\ee
where the subscripts $_\leq 0$ or $_\geq 0$ denote the lower or upper
part.
\end{enumerate}
\ep
{\bf Proof}. The detailed proof of this result may be found in
\cite{BEHH}. Here we just note that,
given the band structure of the two semi-infinite
 matrices $P$ and $Q$, the polynomial $V_1'(x)$ may be uniquely
determined from the relation 
\be
(P-V'_1(Q))\cdot e_0 = 0\ ,\ \ e_0:= (1,0,0,0,...)^t\label{ciccia}
\ee
and its  existence rests upon the assumption that $\gamma(n)\neq 0$.
A similar relation uniquely determines $V_2'(y)$.\\
It may then be shown that all the relation contained in
eq. (\ref{triang}) are satisfied by these polynomials.\\
Conversely, if two polynomials $V_1'$ and $V_2'$ satisfying eq. 
(\ref{triang}) exist, then
\be
[P,Q-V_2(P)]= [P,Q] = [P-V_1(Q),Q]\ .
\ee
But the LHS is upper triangular and the RHS is lower triangular (not
strictly), so that $[P,Q]$ must be diagonal. Q.E.D.\par \vskip 3pt

The structure (\ref{Qdef}), (\ref{Pdef}) of the two matrices $P$ and $Q$ 
means that the four sequences
$\psi_n,\un\psi_n,\phi_n,\un\phi_n$ satisfy both  multiplicative and
 derivative recursion relations
\bea
  x\psi_n &=& \gamma(n) \psi_{n+1} + \sum_{j=0}^{d_2}
\a_j(n)\psi_{n-j},
\\ \ -\frac 1K \frac d{dx} \psi_n &=& \gamma(n-1)\psi_{n-1} +
\sum_{j=0}^{d_1} \beta_j(n+j)\psi_{n+j}\ ,
\\
  y\phi_n &=& \gamma(n) \phi_{n+1} + \sum_{j=0}^{d_1}
\b_j(n)\phi_{n-j},
\\ \ -\frac 1K \frac d{dx} \phi_n &=& \gamma(n-1)\phi_{n-1} +
\sum_{j=0}^{d_2} \a_j(n+j)\phi_{n+j}\ .\label{recs}
\eea
From the finite recursion 
relations satisfied by the  quasi-polynomials $\{\psi_n(x)\}$ and 
$\{\phi_n(y)\}$ follows a set of ``generalized Christoffel--Darboux relations 
\cite{Userkesm, eynardchain}, which imply that the kernels
$\ds{\m{K}^N}_{11}(x,x')$ and 
$\ds{\m{K}^N}_{22}(y',y)$ may be expressed as: 
{\small
\bea
 &&{\m{K}^N}_{11}(x,x')  = 
 {\ds{\le(\m{\un\Phi}^{N\!-\!1} (x'),\Amat^N  \m{\Psi}_{N} (x)\ri)}
\over x'-x} \ , 
\\
 && {\m{K}^N}_{22}(y',y) =
 {\ds{\le(\m{\un\Psi}^{N\!-\!1} (y'), \Bmat^N  \m{\Phi}_{N} (y)\ri)} 
\over y'-y}\  ,
\\
\ \ \Amat^N &:=& \le[
\begin{array}{cccc|c}
0&0&0&0&\!\!-\!\gamma(\!N\!\!-\!1\!)\!\!\cr\hline
\a_{d_2}\!(\!N\!)& \cdots & \a_{2}(\!N\!)& \a_1(\!N\!)& 0\cr
0& \a_{d_2}\!(\!N\!\!+\!1\!) & \cdots & \a_1(\!N\!+\!1\!)& 0\cr
0&0&\a_{d_2}\!(\!N\!\!+\!2\!) &\cdots & 0\cr
0&0&0&\a_{d_2}\!(\!N\!\!+\!d_2\!-\!1\!)&0
\end{array}
\ri]
\\   \ \Bmat^N &:=& \le[
\begin{array}{cccc|c}
0&0&0&0&\!\!-\!\gamma(\!N\!\!-\!1\!)\!\!\cr\hline
\b_{d_1}\!(\!N\!)& \cdots & \b_{2}(\!N\!)& \b_1(\!N\!)& 0\cr
0& \b_{d_1}\!(\!N\!\!+\!1\!) & \cdots & \b_1(\!N\!+\!1\!)& 0\cr
0&0&\b_{d_1}\!(\!N\!\!+\!2\!) &\cdots & 0\cr
0&0&0&\b_{d_1}\!(\!N\!\!+\!d_1\!-\!1\!)&0
\end{array}
\ri]
\label{DCpairing}
\eea
}
where  $\ds{\m{\Psi}_{N} (x)}$ , $\ds{\m{\Phi}_{N} (y)}$,
 $\ds{\m{\un\Psi}^{N-1} (y)}$ and $\ds{\m{\un\Phi}^{N-1} (x)}$ are
the column or row  vectors of dimension $(d_1+1)$ and $(d_2+1)$ 
defined by
\bea
&&\m{\Psi}_{N} (x) = [\psi_{N\!-\!d_2},\dots,\psi_{N}]^t\ ,\ \ 
\m{\Phi}_{N} (y)= [\phi_{N\!-\!d_2},\dots,\phi_{N}]^t,\\
&&\m{\un\Psi}^{N-1} (y) = [\un\psi_{N\!-\!
1},\dots,\un\psi_{N\!+\!d_2\!-\!1}]\ ,\ \  
\m{\un\Phi}^{N-1} (x)=  [\un\phi_{N\!-\! 1},\dots,\un\phi_{N\!+d_1\!-\! 1}].
\eea
The matrices $\ds{\Amat^N}$, $\ds{\Bmat^N}$ entering eqs. \ref{DCpairing} define
two pairings (which we will refer to as the Christoffel-Darboux
pairings) between $\ds{\m{\Psi}_{N}}$ and $\ds{\m{\un\Phi}^{N-1}}$ and
between $\ds{\m{\Phi}_{N}}$ and $\ds{\m{\un\Psi}^{N-1}}$. We call
these  pairs {\em dual windows}.\par
The key observation is that any quasipolynomial $\psi_j(x)$ can be uniquely 
expressed, for any given  $N\ge d_2$, in terms of linear combinations of any 
$d_2+1$ consecutive  basis elements $\psi_{N-d_2},..,\psi_N$ {\em with 
polynomial coefficients}.
 We call this procedure  {\bf folding} of the space onto the {\em
window}  spanned  by
$\m{\Psi}_N =[ \psi_{N-d_2},..,\psi_N]^t$.
This is accomplished by means of the $x$-recursion relations for the
quasipolynomials in eq. (\ref{recs}),
which allow us to express the $(N+1)$st quasipolynomial in terms of
the $d_2+1$ preceding ones, but with coefficients that are polynomials in
$x$.   Iteration of this procedure defines the folding.\par
In matricial form the above can be expressed as follows
\bea
\A_N(x)\m{\Psi}_N(x)&  =
 \ds{\m{\Psi}_{N\!+\!1}}(x)\ ,  \quad N\geq d_1  \ .   \label{bNdef}
\eea
where
\be
\A_N(x):= 
\le[\begin{array}{cccc} 0  & 1 & 0 &\!\!\! \!\!\!\!\!\! \!\!\! 0 \cr
0 & 0 & \ddots &\!\!\! \!\!\!\!\!\! \!\!\!0\cr
0 & 0 & 0 &\!\!\! \!\!\!\!\!\! \!\!\!1\cr
\!\!\frac {-\alpha_{d_2}(N)} {\gamma(N)}\!\! &
\cdots
&\!\! \frac {-\alpha_1(N)}
{\gamma(N)} \!\! &\!\!  \frac{(x-\alpha_0(N))}{\gamma(N)} \!\!  
\end{array}\ri]  
\ee
The matrix $\ds{\A_N}$   is invertible, since its determinant equals
$\a_{d_2}(n)/\g(n)$ and $\a_{d_2}(n)$ can be shown not to vanish as a
consequence of the relation 
\be(Q-V_2'(P))_{\leq 0}=0
\ee
 It will be referred to in the following as a ``ladder'' matrix. 
A completely analogous relation can also be shown for the
quasipolynomials $\phi_n(y)$ (see eq. (\ref{defD2}) below)
 and for the respective Fourier-Laplace
transforms.\par
By means of this folding, one can also express the action of any
operator of finite band size as a matrix polynomial in $x$ of size
$d_2+1$. The most relevant case is the folding of the operator $P=
-\frac 1 K {d\over dx}$. Introducing the notation
\be
\m{\Psi}_N:=[\psi_{N-d_2},...,\psi_N]^t
\ee
 we have
\bea
&& -\frac 1K \frac {d}{dx} \m{\Psi}_N =
 \m{D_1}^N(x)\m{\Psi}_N 
\cr &&:=\le(  
 \stackrel{N}{\gamma}(\m{\A}_{N-1}(x))^{-1} +
\stackrel{N}{\b_0} +
\sum_{j=1}^{d_1}\stackrel{N}{\b_j}\A_{\!N\!+\!j\!-\!1\!}\! (x)
\A_{\!\!N\!+\!j\!-\!2\!}\!(x)\cdots
\A_N\!(x)\ri)\m{\Psi}_N , \label{defD1}
\eea 
where
\bea
 \stackrel{N}\b_j &:=&
{\rm diag}\le[\b_j(N+j-d_2),\b_j(N+j-d_2+1),\dots,\b_j(N+j) \ri]\ ,
\\
&&\qquad(j=0,\dots d_1)
\cr
\stackrel{N}\gamma &:=& {\rm
diag}\le[\gamma(N-1-d_2),\gamma(N-d_2),\dots,\gamma(N+d_2-1) \ri] 
\ .\label{betajN}
\eea

The corresponding statement for the $\phi_n$'s is obtained by
interchanging $x$
and  $y$, $\psi_n$ and $\phi_n$, $d_1$ and $d_2$, $\alpha_j$
and $\beta_j$ etc. 
One  obtains a similarly defined matrix $D_2(y)$ representing the
action of the derivative on the quasipolynomials $\phi_n$'s.
With the notations
\bea
\m{\Phi}_N &:=& [\phi_{N-d_1},...,\phi_N]^t\ ,\\
\stackrel{N}\a_j &:=& 
{\rm diag}\le[\a_j(N+j-d_1),\a_j(N+j-d_1+1),\dots ,\a_j(N+j) \ri]  ,\\
&&\qquad j=0, \dots d_2
\cr  \label{alphajN}
 \stackrel{N}{\gamma} &:=& {\rm diag}\le[\gamma(N-d_1-1), \dots,
\gamma(N-1) \ri] \\
{\B_{N}}(y)  &:=& \le[\begin{array}{cccc} 
  0 & 1& 0 &\!\!\! \!\!\!\!\!\! \!\!\!0 \cr
  0 & 0 & \ddots  &\!\!\! \!\!\!\!\!\! \!\!\!0 \cr
  0 & 0 & 0 &\!\!\! \!\!\!\!\!\! \!\!\! 1 \cr
\!\! \frac {-\b_{d_1}(N)} {\gamma(N)} \!\! &
\cdots
 &\!\! 
\frac { -\b_1(N)} {\gamma(N)} \!\!     &\!\!
\frac{(y-\b_0(N))}{\gamma(N)} 
\end{array}\ri] \ ,  \quad N\geq d_1  
\eea
one finds
\bea 
&& \B_N(y)\m{\Phi}_N(y) = \m{\Phi}_{N\!+\!1}(y)\label{ladder1}\\
&& -\frac 1K \frac{d}{dy}\m{\Phi}_N = 
\m{ D_2}^N(y) \m{\Phi}_N \cr
&&:=\le(
 \stackrel{N}{\gamma} (\m{\B}_{N-1})^{-1}(y) +
\stackrel{N}{\a_0} + \sum_{j=1}^{d_2}\stackrel{N}{\a_j}
{\B_{\!N\!+\!j\!-\!1\!}\!(y)}
{\B_{\!N\!+\!j-\!2\!}\!(y)} \cdots{\B_N}(y)\ri)\m{\Phi}_N. \label{defD2}
\eea 

We can repeat a similar procedure for the
respective Fourier-Laplace transforms.
The relevant definitions and relations are given by the following
formulae
\bea
 \m{\un\A}^{N}(x) &:=& \le[
\begin{array}{cccc} 
  \frac{x\!-\!\a_0\!(N)}{\gamma(N\!-\!1)}&1&0&0\\[5pt]
 \frac{-\a_1\!(N\!+\!1)}{\gamma(N\!-\!1)}&0&
^{\ds{^{\ds{\cdot}}}}\hbox{}
 ^{\ds{\cdot}} \cdot&0\\
 \vdots &0&0&1\\[5pt]
   \frac{-\a_{d_2}\!(N\!+\!d_2)}{\gamma(N\!-\!1)}&0&0&0
\end{array} \ri]\in gl_{d_2+1}[x] 
\label{abarN} \ ;\\ 
 \m{\underline\B}^{N}(y) &:=& \le[
\begin{array}{cccc}
 \frac{y\!-\!\b_0\!(N)}{\gamma(N\!-\!1)}&1&0&0\\[5pt]
 \frac{-\b_1\!(N\!+\!1)}{\gamma(N\!-\!1)}&0&
 ^{\ds{^{\ds{\cdot}}}}\hbox{}
 ^{\ds{\cdot}} \cdot&0\\
 \vdots &0&0& 1\\[5pt]
  \frac{-\b_{d_1}\!(N\!+\!d_1)}{\gamma(N\!-\!1)}&0&0&0
\end{array} \ri]\in gl_{d_1+1}[y] \ ,  \label{bbarN}
\eea
\bea
&& \m{\un\Psi}^{N\!-\!1}=\m{\un\Psi}^{N}\m{\un\A}^{N}(x)\ ,\qquad 
\m{\un\Phi}^{N\!-\!1}=\m{\un\Phi}^{N}\m{\un\B}^{N}(y)\ ,\label{ladder2}\\
&&\frac 1K\frac d{d y}\m{\un\Psi}^{N\!-\!1}(y)  =
\ds{\m{\un\Psi}^{N\!-\!1}(y) {\m{\un D}^N}_2(y)}\ ,
\quad N\ge d_1+1 \ , \label{yDE_PsibarN} \\
&&\frac 1K\frac d{d x}\m{\un\Phi}^{N\!-\!1}(x)  =
\ds{\m{\un\Phi}^{N\!-\!1}(x) {\m{\un D}^{N}}_1(x)} \ , 
\quad N\ge d_2+1 \ , \label{xDE_PhibarN}
\eea
where
\bea
&&{\m{\un D}^N}_2(y) := \ds{(\m{\un\B}^{N})^{-1}
\m{\underline \gamma}^{N-1}  + {\m{\underline\a_0}^{N-1}}
+\sum_{j=1}^{d_2}
\m{\un\B}^{N-1}\m{\un\B}^{N-2}\cdots
\m{\un\B}^{N-j}
{\m{\un\a_j}^{N-1}} }  \ , \\
&&\m{\underline D_1}^N(x)  := \ds{(\m{\un\A}^{N})^{-1}
\m{\underline \gamma}^{N-1}  + {\m{\underline\b_0}^{N-1}}
+\sum_{j=1}^{d_1}
\m{\un\A}^{N-1}\m{\un\A}^{N-2}\cdots
\m{\un\A}^{N-j}
{\m{\underline\b_j}^{N-1}} } \\
&&\m{\un\a_j}^{N\!-\!1}:={\rm diag}(\a_j(N-1),\dots,\a_j(N+d_1-1))\ ,\\  
&&\m{\un\b_j}^{N\!-\!1}:={\rm diag}(\b_j(N-1),\dots,\b_j(N+d_2-1))
\eea
Summarizing, we have  four sequences of linear differential systems
\bea
\begin{array}{l|l}
\hbox {Size } (d_2+1)\times (d_2+1) & \hbox {Size } (d_1+1)\times
(d_1+1) \\
\hline
\vrul 
\ds{
-\frac 1K\frac d{d x}\m{\Psi}_{N}  (x) =  \m
{ D_1}^N(x)\m{\Psi}_{N}
(x) }& \ds{
\frac 1K\frac d{ d y}\m{\underline \Psi}^{N\!-\!1}(y) =
\m{\un\Psi}^{N\!-\!1}(y)\m{\underline
D_2}^N (y)}\\[10pt] 
\ds{\frac 1K\frac d{d x}\m{\underline \Phi}^{N\!-\!1}
(x) =
\m{\un\Phi}^{N\!-\!1}(x)\m{\underline
D_1}^N (x)} &
\ds{-\frac 1K\frac d{d y}\m{\Phi}_{N}  (y) =  \m
{ D_2}^N(y)\m{\Phi}_{N}(y)}
\end{array} \label{foursystem}
\eea
as well as the ladder relations (\ref{bNdef}), (\ref{ladder1}),
(\ref{abarN}), (\ref{ladder2}). 
We have not considered here the deformation  equations, i.e. the
differential equations obtained from  infinitesimal variations of
the coefficients of the potentials $V_1$ and $V_2$ entering the
measure. The complete study of these deformations is carried out in
\cite{BEH}. In particular it is shown there that the resulting
overdetermined system of PDEs is compatible.
Here we will only recall that the mixed system of ODEs and difference
equations is also compatible, as implied by the following:
\bp
\label{shift_dx_compatibility}
The ladder matrices $\ds{\A_N}$ intertwine the
differential systems $D_1$ with different $N$'s, i.e.
\be
 \A_N(x)\le(\frac d{dx} + \m{D_1}^N (x)\ri) =\le(\frac d{dx} +
\m{D_1}^{N+1}(x)\ri)\A_N (x)
\ee
Similar statements hold for the other three sequences of ODEs and
ladder relations.
\ep
\hspace*{3mm}The next proposition explains how the  four sequences of systems in
the Table   are related amongst
themselves by means of the Christoffel--Darboux pairings.
\bp
The following relations are satisfied
\be
\m{\un D_1}^N(x)\Amat^N = \Amat^N \m{D_1}^N(x)\ ;\ \ \m{\un D_2}^N(y)\Bmat^N
= \Bmat^N \m{D_2}^N(y)
\ee
\ep
The spectra of the two matrices
$D_1(x)$ and $\un D_1(x)$ (i.e., their characteristic polynomials) 
coincide, as do the spectra of $D_2(y)$ and $\un D_2(y)$.

A less apparent  spectral duality also holds.
Indeed it is proven in \cite{BEH} that 
\be
\det\le(y\1-\m{D_1}^N(x)\ri) = c\det\le(x\1-\m{D_2}^N (y)\ri)\ ,
\ee
where $c$ is the ratio of the
leading coefficients of the two potentials $V_1$ and $V_2$.
Notice that the two determinants involve square matrices of rank
$d_2+1$ on the LHS and of rank $d_1+1$ on the RHS. In the following
section we give a simple derivation of a ``na\"\i ve''  $N\to\infty$
limit of these results; namely one in which we treat the relevant
recursion matrices as commuting. 
\section{The Abelian case}
\label{tre}
In this section we derive the spectral duality property  
 in a particular limit $N\to
\infty$, $K/N=\mathcal O(1)$.
In such a limit the two matrices $P$ and $Q$, while retaining their
finite band structure, may be taken to  commute because $[P,Q]=-\1/K
 \to 0$.
In addition, we consider only the case in which the coefficients
$\a_j(n)$, $\b_j(n)$, $g(n)$  do not depend on $n$: this is a
stronger requirement which occurs actually only for certain ranges of
the coupling constants. This limit is studied in the literature and 
 is referred to as the ``one-cut case'' or the  ``genus $0$'' case
\cite{eynard,KazakoVDK}.\par
A further simplification that is purely technical is obtained by 
considering the matrices as doubly-infinite, i.e. of size $\Z\times
\Z$ instead of $\N\times\N$.
We will show that the statement of spectral duality in this case 
 reduces to a
classical result in commutative algebra, namely the
computation of the resultant of two Laurent polynomials.

The non-abelian case (i.e. for finite $N$) is detailed in  \cite{BEH} 
and the approach used there may be used to derive the result for the
$N\to\infty$ case. 
 However,  we will present a proof here of a different
nature, which can also be extended to the non-abelian case
\cite{BEHH}.\par
The equations $[P,Q]=-\frac 1 K \1$ in the limit $N\to \infty$,
$K=\mathcal O(N)$, become commutativity equations $[P,Q]=0$. Moreover,
since we are considering finite band matrices and we focus on the
window at $N$, we can replace  the semi--infinite matrices $P,Q$ by
doubly infinite matrices with the same band structure.
For suitable scaling regimes it can be argued on physical grounds that
the sequences $\gamma(n),\ \a_j(n),\ \b_k(n)$ actually do not depend
on $n$ (provided $n=\mathcal O(N)$). It is precisely this very
simple case that  we want to address here.\par
 The  pair of commuting matrices $P$
and $Q$ with the same band structure as before now just become polynomials
in the shift matrix.
All the matrices are taken to be  $\Z\times \Z$ matrices and hence $\Lambda =
[\delta_{i,i+1}]$ is actually invertible, the inverse being just the
transpose $\L^t$. With this in mind we can write 
\bea
&& Q(\L):= \gamma \L  + \alpha_0 + \sum_{i=1}^{d_2} \alpha_i
\L^{-i} \ , \quad \g\neq 0\neq \a_{d_2} \label{354}\\
&& P(\L) := \gamma\L^{-1} + \beta_0 + \sum_{i=1}^{d_1}
\beta_i\L^{i} \ , \quad  \g\neq 0\neq \b_{d_1} \ ,\label{355}
\eea
with  $Q$ and $P$ are viewed as Laurent polynomials in
$\L,\L^{-1}$. It is convenient to introduce an indeterminate
$\lambda$ and represent $Q$ and $P$ as acting on the graded space
\be
Q, P: \C[\lambda,\lambda^{-1}]\to \C[\lambda,\lambda^{-1}]\ ,
\ee
determined by substituting $\Lambda$ by $\lambda$ in the relations
(\ref{354},\ref{355}).
The shift matrix $\L$ is just multiplication by $\lambda$ while $\L^t=
\L^{-1}$ represents multiplication by $\l^{-1}$.
The equivalent of a window is then the linear span of $d_2+1$
consecutive powers of $\l$
\be
\C\{\psi_{N-d_2},...,\psi_{N}\}\leftrightarrow
\C\{\l^{N-d_2},...,\l^N\}\ .
\ee

The folding of the previous sections here reduces to a very simple
expression.
Indeed, folding the graded space $W:=\C[[\lambda]]$ onto the span of
$\l^{N-d_2},..., \l^{N}$ simply means taking the quotient
\be
\C[[\lambda]]\simeq \C[x]\otimes\C[[\l]]\,{\rm mod}\, \langle x
-Q(\l)=0 \rangle\,\, \simeq
\C[x]\{ \l^{N-d_2},...,\l^{N}\}\ . 
\ee
In other words, the power $\l^{N+1}$ can be re-expressed in terms of
the powers $\l^{N-d_2},...,\l^{N}$ using the relation $x-Q(\l)=0$.
The equivalent of the ladder matrix is just the expression of
multiplication by $\l$ in the ``folded'' window $\C[x]\{
\l^{N-d_2},...,\l^{N}\}$. It is defined so as to make  the
following diagram commutative
\be
\begin{diagram}
\C[[\l]] &\rTo^\l & \C[[\l]]\\
\dTo^{\hbox{\small $\langle
\!x\!-\!Q(\!\l\!)\!=\!0 \rangle$}}  & & \dTo_{\hbox{\small $\langle
\!x\!-\!Q(\!\l\!)\!=\!0\rangle$}}\\
\C[x]\{ \l^{N-d_2},...,\l^{N}\}&\rTo^{\A(x)} & \C[x]\{
\l^{N-d_2},...,\l^{N}\}
\end{diagram}
\ee
In principle, $\A$ could depend on $N$, but it is easy to see that in
fact it is represented by the following $N$-independent companion-like matrix 
\be
\A(x) = \le[\begin{array}{cccc}
0&1&\cdots &0\\
0&0&\ddots&\vdots\\
0&0&\cdots &1\\
-\frac {\a_{d_2}}\g&\cdots &-\frac {\a_1}\g& \frac {x-\a_0}\g
\end{array}\ri]
\ee
Similarly, we could define another folding along $P$ by means of the
following diagram
\be
\begin{diagram}
\C[[\l]] &\rTo^\l & \C[[\l]]\\
\dTo^{\hbox{\small $\langle
\!y\!-\!P(\!\l\!)\!=\!0 \rangle$}}  & & \dTo_{\hbox{\small $\langle
\!y\!-\!P(\!\l\!)\!=\!0\rangle$}}\\
\C[y]\{ \l^{N-1},...,\l^{N+d_1-1}\}&\lTo^{\B(y)} & \C[y]\{
\l^{N-1},...,\l^{N+d_1-1}\}
\end{diagram}
\ee
where  $\B$ is given by
\be
\B(y) = \le[\begin{array}{cccc}
0&1&\cdots &0\\
0&0&\ddots&\vdots\\
0&0&\cdots &1\\
-\frac {\b_{d_1}}\g&\cdots &-\frac {\b_1}\g& \frac {y-\b_0}\g
\end{array}\ri]\ .
\ee
In this framework the matrices $D_1(x)$ and $D_2(y)$ are simply
\bea
&& D_1(x):= P(\A(x))= \gamma\A(x)^{-1} +\sum_{j=0}^{d_1}\b_j\A^j(x) \\
&&  D_2(y):= Q(\B(y))= \gamma\B(y)^{-1} + \sum_{j=0}^{d_2} \a_j\B^j(y) \ .
\eea

The previous statement about spectral duality now translates into the
identity
\be
\det(y\1-D_1(x)) \propto \det (x\1-D_2(y))\ .
\ee
We will show that both determinants are
in fact the resultants (w.r.t. $\l$) of the two Laurent polynomials $Q(\l)-x$
and $P(\l)-y$. 
The proof is actually quite standard for polynomials and here we just
adapt it to the situation with Laurent polynomials (see  e.g.  \cite{gelfand}).
This  amounts to studying the following embeddings
\bea
\C\{\l^{N-d_2},...,\l^N\}
\rTo^{\hbox{\scriptsize $P(\l)\!-\!y$}}
 \C\{\l^{N-d_2-1},...,\l^{N+d_1}\}  
\ , \\
 \C\{\l^{N-1},...,\l^{N+d_1-1}\} 
\rTo^{\hbox{\scriptsize $Q(\l)\! -\!x$}}
 \C\{\l^{N-d_2-1},...,\l^{N+d_1}\}  
\ ,
\eea
where $x$ and $y$ are treated as parameters of the embeddings. Let us
denote by $W, \un W, U$ the three vector spaces 
\bea
W &:=& \C\{\l^{N-d_2},...,\l^N\}\ ,  \quad \un W :=
\C\{\l^{N-1},...,\l^{N+d_1-1}\} \ ;
\cr  \ U&:=&
\C\{\l^{N-d_2-1},...,\l^{N+d_1}\}\ .
\eea
The above embedding may be combined into a single map 
\be
\begin{diagram}
W \oplus \un W
&\rTo^{\hbox{\scriptsize $(P(\l)\!-\!y)\!\oplus\! (Q(\l)\! -\!x)$}} U 
\end{diagram}
\ee
The two parts of this map give spaces generically transverse as $x$
and $y$ vary. If
 they are  not transverse for a given pair $(x,y)$, this means that 
\be
\exists \,w\in W,\ \exists \,\un w\in\un W\ \hbox{such that }\ 
 w\neq 0\neq \un w\
, \ (P-y)w = (Q-x)\un
w\,\ \in U \ .
\ee
Taking the quotient of this relation by the relation $Q(\l)-x=0$ or
$P(\l)-y=0$ gives rise to the relation 
\bea
&&(D_1(x)-y) w = 0\label{eigen1}\\
&&(D_2(y)-x)\un w=0\ ,\label{eigen2}
\eea
which means that $y$ is an eigenvalue of $D_1(x)$ and $x$ an
eigenvalue of $D_2(y)$. Conversely if either of the two
 equations (\ref{eigen1}) (\ref{eigen2}) holds, say the first, for 
nonzero vector $w$, this means that there exists $\un w\in \un W$ such that  
\be
(P-y)w = (Q-x)\un w\ .
\ee
Notice that $(Q-x) \un w$ cannot be zero since the map
$P-y:W\to U$ is injective for all $y$ and so $(P-y)w\neq 0$.
The same result follows if we start from  eq. (\ref{eigen2}).
This proves that the embedding is not transverse if and only if $x$ is
 an eigenvalue of $D_2(y)$ which is equivalent to  $y$ being 
an eigenvalue of $D_1(x)$.\par
The condition of transversality amounts to the nonvanishing of the
determinant of the embedding (in any fixed basis).
It is easy to see that such an embedding is represented by the
Sylvester matrix
\be
\le[\begin{array}{ccccccccc}
\g&\b_0\!-\!y & \b_1&\cdots &\cdots &\b_{d_1}& 0 &0&0\\
0&\g&\b_0\!-\!y & \b_1&\cdots &\cdots &\b_{d_1}& 0 &0\\
0&0&\g&\b_0\!-\!y & \b_1&\cdots &\cdots &\b_{d_1}& 0\\
0&0&0&\g&\b_0\!-\!y & \b_1&\cdots &\cdots &\b_{d_1}\\
\hline
\a_{d_2}&\cdots&\a_1&\a_0\!-\!x&\g&0&0&0&0\\
0&\a_{d_2}&\cdots&\a_1&\a_0\!-\!x&\g&0&0&0\\
0&0&\a_{d_2}&\cdots&\a_1&\a_0\!-\!x&\g&0&0\\
0&0&0&\a_{d_2}&\cdots&\a_1&\a_0\!-\!x&\g&0\\
0&0&0&0&\a_{d_2}&\cdots&\a_1&\a_0\!-\!x&\g\\
\end{array}
\ri]
\ee
 of the two Laurent polynomials,  whose
determinant $\Delta(x,y)$ equals the resultant.
A simple counting of degrees and inspection of the highest powers in
$x$ or $y$  shows that 
\be
\a_{d_2}\g^{d_1}\det(y\1-D_1(x)) = \Delta(x,y) =
\b_{d_1}\g^{d_2}\det(x\1-D_2(y))\ .
\ee
which defines the spectral curves as the non-transversality locus of
the embeddings.
The intersection of the two embeddings on
this spectral curve is (generically) one-dimensional and projects to
the eigenvectors of $D_1(x)$ and $D_2(y)$.\par
While this is very simple, and just a reformulation of standard
algebraic results in this abelian setting, 
 a very similar approach can also be used to  prove
spectral duality for the pair $\m{D_1}^N(x)$ and $\m{\un D_2}^{N}(y)$
in the finite $N$ setting, in which the matrices $P$ and $Q$ do not
commute.
A refinement and elaboration on this theme also leads to the other
results of \cite{BEH} in a more elegant and compact form \cite{BEHH},
such as the compatibility of the deformation equations in the coupling
constants of the potentials $V_1$, $V_2$ which, in particular imply
the invariance of the generalized monodromy of the operators $\ds{\pa_x +
\m{D_1}^N(x)}$ and $\ds{\pa_y + \m{D_2}^N(y)}$.
This defines a sort of ``noncommutative resultant'' for finite band
matrices whose properties will be developed in a subsequent
publication.

\end{document}